# Спектры возбуждения-эмиссии люминесценции ИК висмутовых активных центров в волоконных световодах

С. В. Фирстов (1), В. Ф. Хопин (2), И. А. Буфетов (1), Е. Г. Фирстова (1),
А. Н. Гурьянов (2), Е. М. Дианов (1)

((1)-Научный центр волоконной оптики РАН, Москва, Россия)
((2)-Институт химии высокочистых веществ РАН, Н.Новгород, Россия)

**Аннотация:** Впервые получены 3-мерные спектры люминесценции (интенсивность люминесценции в зависимости от длин волн возбуждения и эмиссии) легированных висмутом волоконных световодов различного состава в широком диапазоне длин волн (450-1700 нм). Исследования выполнены для легированных висмутом световодов с сердцевиной из $SiO_2$, $GeO_2$, а также из $SiO_2$ с добавкой Al или P (при комнатной температуре и при температуре жидкого азота). На основании полученых результатов построены схемы энергетических уровней для ИК висмутовых активных центров в световодах с сердцевиной из $SiO_2$ и из $GeO_2$. Обнаружено подобие схем энергетических уровней висмутовых активных центров в световодах указанных составов.

## 1. Введение

Стекла и волоконные световоды, легированные висмутом, являются новыми оптически активными материалами, обладающими широким спектром люминесценции в диапазоне длин волн 1000-1700 нм, время жизни которой в стеклах ряда составов составляет 0.1-1 мс. Интерес к таким материалам связан с возможностью их использования для создания лазеров в указанном диапазоне длин волн и для усиления оптических сигналов в диапазоне 1300-1500 нм для волоконных систем связи следующего поколения. До настоящего времени лазерная генерация и оптическое усиление на волоконных световодах, легированных висмутом, получены в области длин волн 1150-1550 нм [1-8]. Серьезным препятствием на пути совершенствования висмутовых лазеров и усилителей является отсутствие адекватной модели ИК висмутового активного центра (ВАЦ) в оптических средах. Ни одна из обсуждаемых в литературе моделей (их достаточно полный перечень можно найти в обзоре [9]) не удовлетворяет всем имеющимся на сегодняшний день экспериментальным данным (см., напр., обзор [10]). ВАЦ наилучшим образом проявляют свои усилительные и генерационные свойства при условии малой концентрации висмута (обычно менее 0.02 ат.%), и, следовательно, самих ВАЦ. Данное обстоятельство служит существенным препятствием при изучении свойств ВАЦ и требует применения наиболее чувствительных методов исследования. Для исследования ВАЦ нами использовался люминесцентный анализ, который широко применяется и применялся в предшествующих работах в этой области (см., напр., обзор [10] и ссылки в нем). Данная работа отличается тем, что в ней выполнены подробные измерения интенсивности люминесценции ($I_{lum}$) ВАЦ в зависимости от длин волн как эмиссии ($\lambda_{em}$), так и возбуждения ($\lambda_{ex}$), изменяющихся в широком спектральном диапазоне – от 450 до 1700 нм. Это позволило построить наглядные контурные графики зависимости $I_{lum}(\lambda_{em}, \lambda_{ex})$ для анализа структуры энергетических уровней ВАЦ. Таким образом исследовались люминесцентные свойства ВАЦ в световодах с сердцевиной наиболее простого состава. А именно: с сердцевиной из $\nu$-$SiO_2$, легированного $Bi_2O_3$ без каких-либо еще легирующих добавок, и с сердцевиной из $\nu$-$GeO_2$, также легированного $Bi_2O_3$. В таких световодах можно ожидать результатов, более "прозрачных" для последующей интерпретации. Получены также спектры $I_{lum}(\lambda_{em}, \lambda_{ex})$ для висмутовых волоконных световодов с сердцевинами из кварцевого стекла, легированного алюминием и фосфором. Все эти данные представляют большой интерес, поскольку именно на висмутовых световодах с добавками алюминия, германия и фосфора наблюдалось оптическое усиление и лазерная генерация в ИК диапазоне в большинстве работ, опубликованных до настоящего времени. Предварительные сообщения по результатам исследования световодов с сердцевиной состава $SiO_2$-$Bi_2O_3$ опубликованы в работах [11, 12].

## 2. Экспериментальные образцы, методы измерений.

Таблица 1. Обозначения, состав сердцевины и метод изготовления исследованных волоконных световодов

|   | Обозначение | Состав стекла сердцевины | Метод изготовления |
|---|---|---|---|
| 1 | SBi | $100SiO_2$+Bi | powder-in tube |
| 2 | GBi | $100GeO_2$+Bi | MCVD |
| 3 | ASBi | $3Al_2O_3$+$97SiO_2$+Bi | MCVD |
| 4 | PSBi | $10P_2O_5$+$90SiO_2$+Bi | MCVD |

Для измерения спектров люминесценции в настоящей работе были выбраны 4 световода с различным составом сердцевины (см. Таблица 1). Концентрация висмута в сердцевинах световодов не превышала порога чувствительности нашей измерительной аппаратуры (0.02 at%), почему она и не указана в таблице. Относительную концентрацию BAC в стекле сердцевины световодов можно оценить по измеренным спектрам поглощения (см. Рис. 1 - Рис. 4). Все световоды имели внешний диаметр 125 мкм. Исследовались как одномодовые (на длине волны 1.2 мкм), так и многомодовые световоды. В условиях проведенных измерений различий между ними не обнаружено.

Световод SBi был изготовлен по технологии powder-in-tube [11]. Его сердцевина была окружена отражающей оболочкой из кварцевого стекла, показатель преломления которого был понижен за счет легирования фтором. Остальные световоды были изготовлены по MCVD технологии, причем все легирующие добавки вводились из газовой фазы. Следует отметить, что при изготовлении преформы по MCVD технологии на внутреннюю поверхность опорной трубы (Heraeus F300) сначала наносился дополнительный слой высокочистого кварцевого стекла, легированного фтором и фосфором (с концентрацией оксида фосфора $C_{P2O5} \leq 1mol\%$, концентрация фтора выбиралась из условия компенсации возрастания показателя преломления кварцевого стекла за счет легирования фосфором). Целью данной операции было снижение потерь излучения при его распространении по сердцевине одномодового световода. После этого уже наносились слои стекла, соответствующие сердцевине световода. Следует отметить, что, поскольку на границе сердцевина – отражающая оболочка в наших световодах имеют место значительные изменения состава (например, наличие фосфора в составе оболочки и его отсутствие в сердцевине световодов 2 и 3, Таблица 1), то во время вытяжки световода процессы диффузии могут модифицировать состав стекла в приграничной области.

Все световоды вытягивались из заготовок в одинаковых условиях, которые включали нагрев до температуры около 2000°C с последующим быстрым охлаждением световода в процессе вытяжки до температур ниже температуры стеклования за время порядка нескольких десятых секунды.

На оптические свойства BAC в стекле сердцевины световода влияет состав стекла, его окислительно-восстановительные свойства, наличие окислительной или восстановительной атмосферы на различных технологических этапах изготовления стекла сердцевины, температурные режимы, скорость охлаждения стекла (особенно процесс вытяжки световода) и т.п. Например, в некоторых случаях люминесцентные свойства преформы и световода, вытянутого из этой же преформы, существенно отличались [13, 14]. Если же технологические условия изготовления световодов отличающихся составов близки (что имеет место в данной работе), то можно говорить, что их оптические свойства определяются, главным образом, составом сердцевины световода. Именно с такой точки зрения в настоящей работе обсуждаются различные ВАС, связанные с тем или иным составом сердцевины.

Оптические потери в световодах измерялись обычным способом путем сравнения пропускания излучения отрезками световода различной длины. Люминесценция сердцевины регистрировалась через боковую поверхность световода (для исключения влияния явления перепоглощения). Для возбуждения люминесценции световода на различных длинах волн использовался источник излучения суперконтинуума (supercontinuum light source) SC450 фирмы Fianium. Узкополосное излучение ($\Delta\lambda=3$нм) выделялось из широкого спектра с помощью акустооптического фильтра и вводилось в сердцевину световода. Спектры люминесценции регистрировались спектроанализатором HP в диапазоне длин волн 875нм<$\lambda_{em}$<1700 нм и спектрометром SP2000 (Ocean Optics) для 450 нм<$\lambda_{em}$<875 нм. Таким образом, были получены спектры люминесценции при изменении $\lambda_{ex}$ в диапазоне 450-1700 нм с шагом 10 нм. Величина шага изменения $\lambda_{ex}$ определяла точность измерения положения максимумов зависимости $I_{lum}(\lambda_{em}, \lambda_{ex})$ по длинам волн. Полученные спектры люминесценции были исправлены на спектральную чувствительность канала регистрации и нормировались на введенную в световод мощность излучения накачки. Измерения проводились при комнатной температуре (RT) и при температуре кипения жидкого азота (LNT).

### 3. Экспериментальные результаты

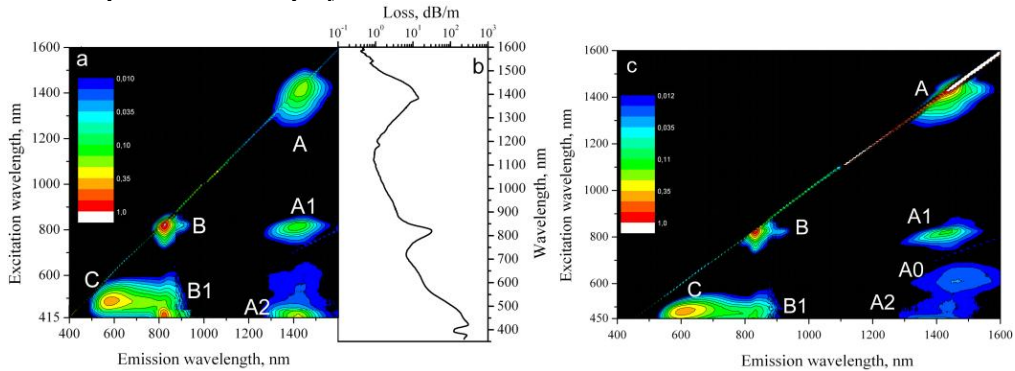

Рис. 1 Зависимость интенсивности люминесценции ВАС от длины волны люминесценции и длины волны возбуждения для SBi световода при T=300 K (a) и T=77K (c); спектр оптических потерь в SBi световоде (b).

Таблица 2. Основные максимумы люминесценции ВАС SBi световода: обозначения, длины волн возбуждения и эмиссии при T=300K и 77K (см.Рис. 1).

| Luminescence maximum | $\lambda_{ex}^{max}$, nm | | $\lambda_{em}^{max}$, nm | |
|---|---|---|---|---|
| | RT | LNT | RT | LNT |
| A | 1415 | 1425 | 1430 | 1435 |
| A1 | 823 | 821 | 1430 | 1430 |
| A2 | 420 | <450 | 1430 | ≤1430 |
| A0 | — | 620 | — | 1480 |
| B | 823 | 823 | 827 | 833 |
| B1 | 420 | <450 | 827 | 830 |
| C | 480 | 480 | 590 | 605 |

Измеренные спектры люминесценции в широкой спектральной области в сочетании с пошаговым изменением длины волны возбуждающего излучения позволили построить зависимости $I_{lum}(\lambda_{em}, \lambda_{ex})$ для всех исследованных световодов. Эти зависимости представлены на Рис. 1-Рис. 4 (a и c) в виде контурных графиков, которые представляют собой сочетание спектров люминесценции (вдоль горизонтальной оси) и спектров возбуждения люминесценции (вдоль вертикальной оси). На этих же рисунках приведены спектры оптических потерь в световодах при RT. Измерения для световодов, легированных висмутом, выполнены как при RT, так и при LNT.

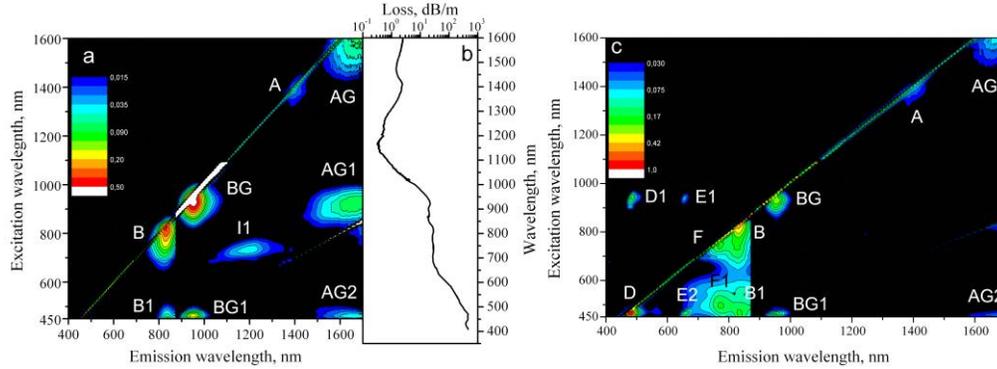

Рис. 2 Зависимость интенсивности люминесценции BAC от длины волны люминесценции и длины волны возбуждения для GBi световода при T=300 K (a) и T=77K (c); спектр оптических потерь в GBi световоде (b).

Таблица 3. Основные максимумы люминесценции BAC GBi световода: обозначения, длины волн возбуждения и эмиссии при T=300K и 77K (см. Рис. 2).

| Luminescence maximum | $\lambda_{ex}^{max}$, nm | | $\lambda_{em}^{max}$, nm | |
|---|---|---|---|---|
| | RT | LNT | RT | LNT |
| A | 1390 | 1390 | 1410 (1410) | 1435 |
| B | 815 | 820 | 830 | 835 |
| B1 | 450 | 450 | 830 | 835 |
| AG | >1600 | >1600 | ≈1670 | ≈1670 |
| AG1 | 925 | — | 1670 | — |
| AG2 | 463 | 463 | 1670 | 1670 |
| BG | 925 | 925 | 955 | 955 |
| BG1 | 463 | 465 | 955 | 955 |
| D | — | 463 | — | 480 |
| D1 | — | 940 | — | 482 |
| E1 | — | 940 | — | 655 |
| E2 | — | 463 | — | 655 |
| F | — | 750 | — | 775 |
| F1 | — | 500 | — | 775 |
| I1 | 735 | — | 1205 | — |

На приведенных графиках зависимостей $I_{lum}(\lambda_{em}, \lambda_{ex})$ отношение максимального и минимального значений интенсивностей люминесценции, отображенных на каждом графике, выбрано ~100. Это ограничивает объем информации, отображаемой на каждом графике, а именно это касается областей с низкими значениями интенсивности люминесценции. Но без такого ограничения трехмерные графики люминесценции получаются практически не читаемыми. Поэтому следует учесть, что на приведенных 3-мерных графиках отображены только основные, наиболее яркие пики наблюдаемой люминесценции.

Обозначения пиков люминесценции, используемые на графиках $I_{lum}(\lambda_{em}, \lambda_{ex})$, и соответствующие максимумам этих пиков длины волн возбуждения и эмиссии, представлены в таблицах 2-5 для RT и для LNT. В дальнейшем для обозначения максимума люминесценции будут указываться его буквенное обозначение в таблице (или на рисунке) и соответствующие ему длины волн возбуждения и эмиссии, например A($\lambda_{ex}^{max}$, $\lambda_{em}^{max}$).

Спектры люминесценции всех световодов представляют собой совокупность пиков, обычно сложной формы и часто налагающихся друг на друга. Спектры световодов различного состава значительно отличаются между собой. Наблюдаемые различия в спектрах люминесценции одного и того же световода при RT и LNT значительно меньше. Ширина пиков люминесценции на графиках $I_{lum}(\lambda_{em}, \lambda_{ex})$ ~100 нм или больше, что объясняется, по-видимому, неоднородным уширением в стеклянной матрице. На всех приведенных графиках $I_{lum}(\lambda_{em}, \lambda_{ex})$ наблюдается линия, проходящая по диагонали $\lambda_{em} = \lambda_{ex}$, которая соответствует рассеянному излучению возбуждения. На всех графиках присутствует также излучение второго порядка дифракции рассеянного излучения накачки в виде участка линии $\lambda_{em} = 2\lambda_{ex}$.

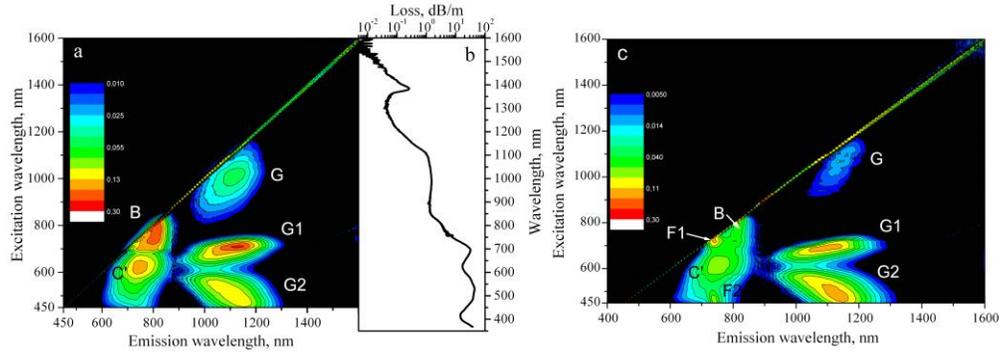

Рис. 3 Зависимость интенсивности люминесценции BAC от длины волны люминесценции и длины волны возбуждения для ASBi световода при T=300 K (a) и T=77K (c); спектр оптических потерь в ASBi световоде (b).

Таблица 4. Обозначения, длины волн возбуждения и люминесценции основных максимумов люминесценции BAC ASBi световода

| Luminescence maximum | $\lambda_{ex}^{max}$, nm | | $\lambda_{em}^{max}$, nm | |
|---|---|---|---|---|
| | RT | LNT | RT | LNT |
| G | 1010 | 1080 | 1120 | 1150 |
| G1 | 705 | 688 | 1130 | 1100 |
| G2 | 510 | 490 | 1100 | 1130 |
| B | 770 | 800 | 815 | 820 |
| F | — | 720 | — | 740 |
| F1 | — | 470 | — | 740 |
| C′ | 620 | 610 | 745 | 745 |

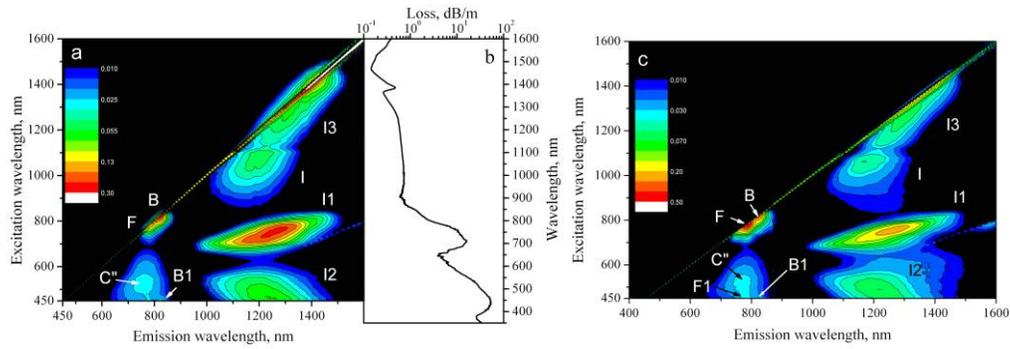

Рис. 4 Зависимость интенсивности люминесценции BAC от длины волны люминесценции и длины волны возбуждения для PSBi световода при T=300 K (a) и T=77K (c); спектр оптических потерь в PSBi световоде (b).

Таблица 5. Обозначения, длины волн возбуждения и люминесценции основных максимумов люминесценции BAC PSBi световода.

| Luminescence maximum | $\lambda_{ex}^{max}$, nm | | $\lambda_{em}^{max}$, nm | |
|---|---|---|---|---|
| | RT | LNT | RT | LNT |
| I | 1065 | 1056 | 1195 | 1180 |
| I1 | 750 | 753 | 1250 | 1260 |
| I2 | ≈460 | ≈460 | ≈1245 | ≈1240 |
| I3 | (1266-1427) | (1241-1409) | (1283-1429) | (1282-1425) |
| B | 815 | 815 | 824 | 825 |
| B1 | 450 | 450 | 825 | 825 |
| F | 785 | 785 | 790 | 790 |
| F1 | — | ≤450 | — | 790 |
| C″ | 523 | 520 | 760 | 772 |

## 4. Обсуждение экспериментальных результатов

*Световод из чистого кварцевого стекла, легированный висмутом.*

Наиболее простой вид трехмерный спектр возбуждения-эмиссии люминесценции $I_{lum}(\lambda_{em}, \lambda_{ex})$ имеет для SBi световода. На спектре Рис. 1a наблюдаются 6 основных максимумов люминесценции в видимой и ИК-области спектра при RT: A, A1, A2, B1, B2 и C (см. Таблица 2). Пики A(1415 нм, 1430 нм) и B(823 нм, 827 нм) отличаются малым стоксовым сдвигом, который значительно меньше их ширины. Они практически лежат на диагонали $\lambda_{ex}^{max} = \lambda_{em}^{max}$. Пики же A1, A2, B1 и C, напротив, характеризуются значительным стоксовым сдвигом между длиной волны возбуждения люминесценции и наблюдаемой длиной волны эмиссии люминесценции. Все упоминавшиеся выше пики с литерами A и B дают люминесценцию в полосах с $\lambda_{em}^{max}$ =1430 нм и $\lambda_{em}^{max}$ =827 нм и имеют попарно одинаковые (за исключением пика A) длины волн возбуждения люминесценции (B и A1 – 823 нм, B1 и A2 – 415 нм или несколько короче).

Основные максимумы зависимости $I_{lum}(\lambda_{em}, \lambda_{ex})$ для SBi световода при RT наблюдаются и при LNT (Рис. 1c). При снижении температуры значительно уменьшается антистоксова часть пиков люминесценции A и B, происходит сужение полос возбуждения люминесценции, а ширина полос эмиссии люминесценции остается практически неизменной. При сужении основных полос возбуждения становятся более ясно видны полосы слабой люминесценции: вблизи пика B(820нм, 830нм) отчетливо наблюдаются более слабые максимумы B′(820нм,910нм) и B″(760нм, 830нм).

Один из них (B′) расположен в области более длинных волн эмиссии люминесценции, чем основной максимум (B), но с одинаковыми длинами волн возбуждения, а другой максимум (B″) – в области более коротких длин волн возбуждения с одинаковой длиной волны эмиссии люминесценции (по отношению к B). Менее отчетливо максимумы B′ и B″ наблюдаются и при RT. При этом максимум B имеет вид сглаженного креста. При LNT появляется новый сравнительно слабый широкий пик ИК люминесценции с максимумом A0(620нм; 1480нм), который практически не наблюдался при RT. Пик A2 становится значительно слабее и, по-видимому, смещается по длинам волн возбуждения в коротковолновую сторону.

Отдельно следует остановиться на пике красной люминесценции C (максимум эмиссии люминесценции лежит около 600 нм). Он значительно отличается от пиков A и B тем, что с ним по длине волны возбуждения не совпадают никакие пики в ИК области спектра. Это указывает на отсутствие связи между источником этой люминесценции и ИК висмутовыми активными центрами в SBi световоде.

Ранее уже высказывалось предположение о принадлежности пика красной люминесценции в SBi световоде ионам $Bi^{2+}$ [11] на основании имеющихся данных о люминесценции $Bi^{2+}$ в кристаллах [15]. Дополнительные аргументы в пользу этого предположения дает сравнение полученных нами спектров возбуждения люминесценции SBi световода на длине волны 600 нм со спектрами возбуждения люминесценции ионов $Bi^{2+}$ в кристаллах [17]. Спектр возбуждения красной люминесценции SBi световода приведен на Рис. 5. На том же графике представлены спектры возбуждения люминесценции ионов $Bi^{2+}$ в кристаллах $SrB_6O_{10}:Bi^{2+}$ и $SrB_4O_7:Bi^{2+}$ [17]. Максимум спектра люминесценции в кристалле $SrB_6O_{10}:Bi^{2+}$ приходится на 660нм, в кристалле $SrB_4O_7:Bi^{2+}$ на 588 нм. Спектры возбуждения ионов $Bi^{2+}$ в кристаллах в диапазоне 225-600 нм состоят из трех пиков, которые соответствуют

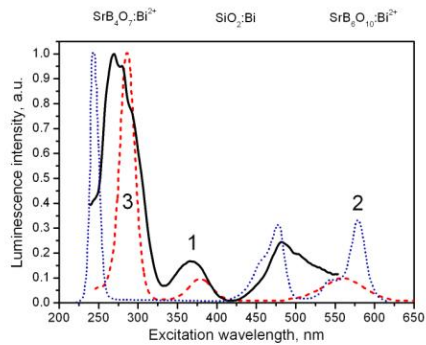

Рис. 5 Спектры возбуждения красной люминесценции: SBi световод (1), $SrB_6O_{10}:Bi^{2+}$ (2) и $SrB_4O_7:Bi^{2+}$ (3)[17].

переходам в ионе $Bi^{2+}$: $^2P_{1/2} \to {}^2S_{1/2}$, $^2P_{1/2} \to {}^2P_{3/2}(2)$, $^2P_{1/2} \to {}^2P_{3/2}(1)$ (в порядке убывания энергии перехода). Люминесценция наблюдается на переходе $^2P_{3/2}(1) \to {}^2P_{1/2}$.

Красная люминесценция в SBi световоде имеет максимум на длине волны 590 нм. Спектр возбуждения этой люминесценции также имеет 3 пика. Причем следует отметить, что различия в положении пиков люминесценции и возбуждения для двух кристаллов существенно больше, чем различия в положении этих максимумов между спектрами кристалла $SrB_6O_{10}:Bi^{2+}$ и SBi световода. Наблюдаемое качественное и в существенной мере количественное подобие спектров люминесценции и возбуждения является подтверждением принадлежности пика люминесценции C(480нм,590нм) ионам $Bi^{2+}$ в сетке кварцевого стекла. И, следовательно, ионы $Bi^{2+}$ не связаны, по-видимому, с ИК висмутовыми активными центрами.

Измеренный спектр поглощения SBi световода представлен на Рис. 1b. В спектре присутствуют полосы поглощения на 420 нм, 830 нм, 620 нм, 1400 нм, плечо на 480 нм. В коротковолновой части спектра наблюдаются полоса поглощения на 390 нм и узкий провал на длине волны 400 нм [11]. Сравнивая графики возбуждения-эмиссии люминесценции (Рис.1a) и оптических потерь (Рис.1b), можно констатировать, что полосы возбуждения основных пиков люминесценции на Рис. 1a совпадают с полосами поглощения на Рис. 1b.

Полагая, что к ВАЦ в SBi световодах относится люминесценция в пиках A, A1, A2, B, B1, можно определить положение первых трех энергетических уровней ВАЦ в кварцевом стекле (ВАЦ-$SiO_2$, см. схему уровней на рис. Рис. 6a; комментарии к переходам 1S и 2S на этом рисунке – см. ниже).

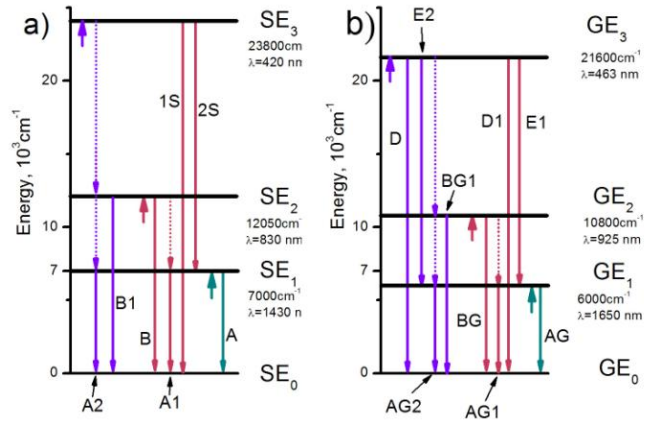

Рис. 6. Схемы энергетических уровней ВАЦ-Si (a) и ВАЦ-Ge (b). На схемах сплошными линиями со стрелками, направленными вниз, обозначены наблюдаемые оптические переходы при RT и LNT. Пунктиром обозначены переходы, люминесценция которых нами не наблюдалась. Короткими стрелками, направленными вверх, обозначена энергия квантов излучения возбуждения. Справа от каждой короткой стрелки обозначены соответствующие такой накачке переходы между уровнями.

*Германатный световод, легированный висмутом*

По сравнению с SBi световодом, GBi световод имеет более сложный вид спектра люминесценции (см. Рис. 2). В сердцевину световода GBi специально не вводились никакие дополнительные легирующие добавки. Но в спектре на Рис. 2а наблюдаются пики A, B, и B1, которые по длинам волн очень близки к одноименным пикам люминесценции SBi световода на Рис. 1а, хотя и сравнительно менее яркие. В GBi световоде наблюдается также набор пиков AG, AG1, AG2, BG и BG1, которые расположены на графике Рис. 2а геометрически подобно пикам A, A1, A2, B и B1 на спектре SBi световода (Рис. 1а), но с небольшим смещением в длинноволновую сторону как по длинам волн возбуждения, так и по длинам волн эмиссии. Таким образом, в спектре люминесценции GBi световода присутствуют как линии люминесценции ВАЦ-$SiO_2$, так и линии люминесценции другой разновидности висмутовых активных центров (пики AG, AG1, AG2, BG и BG1 на Рис. 2), имеющих несколько отличную структуру энергетических уровней, представленную на Рис. 6b. Так как появление таких висмутовых активных центров, несомненно, связано с составом сердцевины световода, то в дальнейшем будем его обозначать как ВАЦ-$GeO_2$. Отметим, что в GBi световоде не наблюдается пик C, соответствующий люминесценции иона $Bi^{2+}$.

Присутствие ВАЦ-$SiO_2$ в GBi световоде можно следующим образом. Сердцевина GBi световода граничит со стеклом отражающей оболочки, состоящей в основном из $SiO_2$. Поле моды волоконного световода проникает в эту оболочку. Атомы висмута могут диффундировать в области около границы сердцевины, состоящие в основном из $SiO_2$, а атомы кремния - из оболочки в сердцевину. Поэтому вполне вероятно, что мы можем наблюдать в GBi световоде, кроме BAC-$GeO_2$, еще и BAC-$SiO_2$. Отметим, что в алюмосиликатном волоконном световоде, легированном висмутом, ранее были обнаружены как висмутовые активные центры, ассоциированные с кремнием, так и ассоциированные с алюминием [18].

В отличие от SBi световода, спектр поглощения GBi не имеет изолированных отдельно расположенных пиков (Рис. 2b). Спектр поглощения в ИК области содержит широкие взаимно перекрывающиеся полосы с максимумами около ≈1600-1650нм. Также наблюдается сложный набор полос в диапазоне 700-1100 нм. В видимой области присутствует интенсивная полоса поглощения на ≈450 нм и "провал" (минимум поглощения) около 425 нм (в SBi световодах аналогичный, по-видимому, минимум наблюдается на ≈400 нм, см. Рис. 1b).

При измерении люминесценции при LNT в спектрах GBi световода появляется дополнительно ряд узких максимумов, расположенных в видимом спектральном диапазоне (максимумы D, D1, E1, E2, F1, F2, см. Рис. 2c, Таблица 3). Так, при LNT невооруженным глазом можно наблюдать яркое синее свечение GBi световода при вводе в сердцевину излучения накачки на длине волны 925 нм мощностью около 1 мВт. Исследование спектров люминесценции показывает, что при возбуждении на длинах волн около 925 нм наблюдаются даже две полосы антистоксовой люминесценции: синяя D1(940, 482) и более слабая красная E1 (940, 655) (Рис. 2c). Такая же пара полос люминесценции возникает при возбуждении на длине волны 460 нм (пики D и E2). Максимум D имеет малый стоксов сдвиг, аналогично пикам A, B, AG и BG, что указывает на его принадлежность переходу $GE_3 \rightarrow GE_0$ в BAC-$GeO_2$ (Рис. 6b). Рядом с пиком люминесценции B при LNT появляется новый пик F, также с малой величиной стоксова сдвига. Кроме того, люминесценция с максимумом на той же длине волны, что и пик F, возникает при возбуждении на 500нм (F1). Возникновение новых линий люминесценции GBi световодов при LNT по сравнению с измерениями при RT может быть объяснено как результат уменьшения вероятности безызлучательной релаксации возбужденных энергетических уровней BAC с уменьшением температуры (уменьшение роли температурного тушения). Данные по люминесценции GBi световодов при LNT находятся в полном соответствии со схемой энергетических уровней BAC-$GeO_2$ (Рис. 6b).

Следует отметить, что наблюдение синей антистоксовой люминесценции при возбуждении GBi световода в ИК диапазоне на длине волны 940 нм, согласно схеме уровней на Рис. 6b, имеет место вследствие поглощения из возбужденного состояния – переходов с уровня $EG_2$ на уровень $EG_3$ под действием того же излучения возбуждения с длиной волны 940 нм. Это возможно при выполнении условия $2EG_2 \approx EG_3$ (см. Рис. 6b, равенство выполняется с точностью до уширения линий). Аналогичное условие $2SG_2 \approx SG_3$ выполняется и в SBi световоде для BAC-$SiO_2$. Но соответствующее радиационному распаду уровня $SG_3$ (Рис. 6a) фиолетовое излучение с длиной волны около 410 нм в наших обычных экспериментах по измерению люминесценции (как они описаны в разделе 2) зафиксировано не было.

Для повышения чувствительности схемы измерения люминесценции мы использовали излучение лазерного диода с длиной волны 803 нм, что позволило увеличить примерно на порядок мощность возбуждения люминесценции в SBi световоде. Спектры наблюдаемой в этом случае антистоксовой люминесценции представлены на Рис. 7. Наблюдаемые при LNT полосы 1S (420нм) и 2S (580 нм) по длинам волн соответствуют переходам $SG_3 \rightarrow SG_0$ и $SG_3 \rightarrow SG_1$ (Рис. 6a). Их сравнительно низкая интенсивность (относительно аналогичных переходов в GBi световоде) может объясняться менее точным выполнением резонансного условия $2SG_2 \approx SG_3$ в BAC-$SiO_2$ по сравнению с BAC-$GeO_2$.

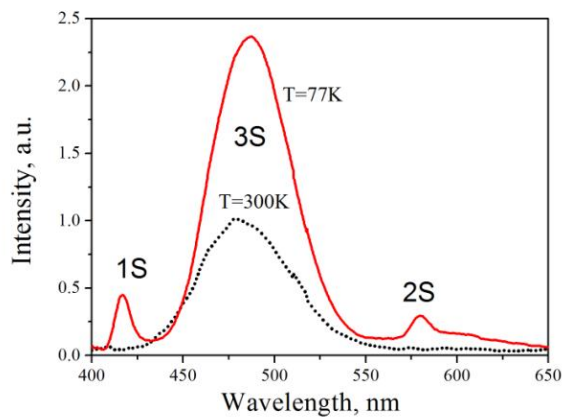

Рис. 7. Антистоксова люминесценция SBi световода при возбуждении лазерным диодом на длине волны 803 нм при RT и при LNT.

Наблюдаемая в таких условиях при RT и LNT полоса синей люминесценции 3S, как и полоса C (Рис. 1а), не соответствует никаким переходам в схеме энергетических уровней BAC-SiO$_2$ (Рис. 6а). Выяснение ее природы является предметом дополнительного исследования.

*Алюмосиликатный и фосфоросиликатный световоды, легированные висмутом*
Картина люминесценции висмутовых световодов качественно изменяется при введении в состав стекла сердцевины дополнительных легирующих примесей, таких как оксид алюминия или оксид фосфора. Спектры поглощения и трехмерные контурные графики люминесценции для ASBi и PSBi световодов приведены на Рис. 3 и Рис. 4 (см. также Таблица 4 и Таблица 5). В этих световодах наблюдаемая люминесценция соответствует висмутовым активным центрам со значительно отличающимися свойствами по сравнению с BAC-SiO$_2$ и BAC-GeO$_2$.

Для висмутовых активных центров в ASBi световодах (BAC-Al) и в PSBi световодах (BAC-P) характерно наличие гораздо более широких полос люминесценции, чем для BAC-SiO$_2$ и BAC-GeO$_2$. BAC-Al и BAC-P отличаются также значительно более сильной зависимостью спектра люминесценции от длины волны возбуждения для некоторых полос люминесценции, о чем свидетельствует значительный наклон этих полос люминесценции на Рис. 3 и Рис. 4 по отношению к оси длин волн эмиссии (таких, как G, G2 для ASBi и I3, I2 для PSBi световода). Подобное поведение полос люминесценции регистрировалось ранее для некоторых стекол и световодов, легированных висмутом [19, 20].

Измерение спектров возбуждения пиков люминесценции C′ в ASBi световоде (Рис. 3) и C″ в PSBi световоде (Рис. 4) показали, что эти спектры качественно (по количеству полос возбуждения) и в некоторой степени количественно (по соотношению амплитуд пиков возбуждения) подобны таким же спектрам для пика C SBi световода и люминесценции кристаллов, содержащих ионы висмута в состоянии B$^{2+}$ (см. Рис. 5). На основании этого можно заключить, что пики люминесценции C′ в ASBi световоде и C″ в PSBi световоде, по-видимому, также принадлежат ионам Bi$^{2+}$.

Полоса люминесценции B в ASBi световоде при RT (Рис. 3а) представляет собой переналожение нескольких полос. Среди них на основании измерений люминесценции при LNT можно выделить пики B и F (Рис. 3с).

Присутствие в спектрах люминесценции ASBi световода пика B (Рис. 3b) и в спектрах PSBi световода пиков B и B1 (Рис. 4), близких по длинам волн к одноименным пикам на Рис. 1, указывает на присутствие в этих световодах BAC-SiO$_2$.

В отличие от SBi и GBi световодов, спектры возбуждения-люминесции для ASBi и PSBi световодов не позволяют очевидным образом построить схему уровней соответствующего висмутового активного центра, как это было сделано для SBi и GBi световодов (Рис. 6). Значительные стоксовы сдвиги частоты большинства линий свидетельствуют о существенном влиянии в данном случае электрон-фононного взаимодействия.

Сравнение пиков люминесценции PSBi и GBi световодов позволяет констатировать близость по длинам волн пиков F, F1 и I1 PSBi световода (Рис. 4) с одноименными пиками люминесценции в GBi световоде (Рис. 2). Полагая, что некоторое отличие положений максимумов люминесценции на этих спектрах может быть следствием влияния матрицы стекла, а также учитывая, что спектральное разрешение наших экспериментов находится на уровне 10 нм, можно сделать вывод, что пики люминесценции F, F1 и I1, наблюдаемые в GBi световоде могут быть обусловлены формированием в этих световодах BAC-P (как указывалось ранее, отражающая оболочка световодов GBi содержит ~1моль% фосфора, и формирование таких центров возможно).

Что же касается ASBi световода, то в нем положение пиков F и F1 (Рис. 3b) существенно отличается от одноименных пиков в PSBi световоде и, следовательно, они не могут быть объяснены как результат присутствия фосфора в алюмосиликатном световоде.

Таким образом, к BAC-Al на Рис. 3 можно отнести максимумы люминесценции G, G1 и G2, F и F1. В спектре люминесценции PSBi световода к BAC-P относятся пики люминесценции I, I1, I2, I3, а также F и F1 (Рис. 4).

## 7. Заключение

В настоящей работе впервые исследованы люминесцентные свойства ряда волоконных световодов с сердцевиной различного состава, легированных висмутом в широком диапазоне длин волн возбуждения и эмиссии люминесценции (450-1700нм).

Все результаты настоящей работы и ряда предшествующих работ [10, 21] по исследованию активных центров в висмутовых волоконных световодах на основе чистого кварцевого стекла, германатного стекла и кварцевого стекла, дополнительно легированного алюминием и фосфором, показывают, что во всех таких световодах присутствуют висмутовые активные центры, свойства которых существенно зависят от состава сердцевины световода (при условии, что уровень легирования висмутом и технологические условия изготовления световода не изменяются). BAC-SiO$_2$ в некоторой степени присутствуют и в GBi, ASBi и PSBi световодах, где предполагается наличие SiO$_2$ в сердцевине (хотя бы небольшой концентрации, как в случае GBi световода). Активные центры типа BAC-Ge обнаружены только в GBi световоде. BAC-P обнаружены в PSBi и GBi световодах, изготовленных методом MCVD, что связано с использованием в отражающей оболочке слоев SiO$_2$, легированных фосфором.

Определено положение низколежащих энергетических уровней висмутовых активных центров, ассоциированных с кремнием и с германием. Показано, что схемы энергетических уровней BAC-SiO$_2$ и BAC-GeO$_2$ подобны друг другу, с более низкими значениями энергии уровней BAC-GeO$_2$: каждый уровень лежит на 10-16% ниже по энергии по сравнению с соответствующим ему уровнем в BAC-SiO$_2$.